\RequirePackage{fix-cm}
\documentclass[twocolumn,epjc3]{svjour3}  
\smartqed  
\RequirePackage{graphicx}
\journalname{Eur. Phys. J. C}
\usepackage{amsmath}
\begin{document}

\title{Black Holes and Warp Drive}


\author{Remo Garattini\thanksref{e1,addr1}
        \and
        Kirill Zatrimaylov\thanksref{e2,addr1} 
}

\thankstext{e1}{e-mail: remo.garattini@unibg.it}
\thankstext{e2}{e-mail: kirill.zatrimaylov@sns.it,\\
kirill.zatrimaylov@guest.unibg.it}

\institute{Università degli Studi di Bergamo, Dipartimento di Ingegneria e Scienze Applicate, Viale Marconi 5, 24044, Dalmine (Bergamo), Italy \label{addr1}}

\date{Received: date / Accepted: date}

\maketitle

\begin{abstract}
We study the generalizations of the original Alcubierre warp drive metric to the case of curved spacetime background. We find that the presence of a horizon is essential when one moves from spherical coordinates to Cartesian coordinates in order to avoid additional singularities. For the specific case of Schwarzschild black hole, the horizon would be effectively absent for the observers inside the warp bubble, implying that warp drives may provide a safe route to cross horizons. Moreover, we discover that the black hole's gravitational field can decrease the amount of negative energy required to sustain a warp drive, which may be instrumental for creating microscopic warp drives in lab experiments. A BEC model is also introduced to propose possible test in the Analogue Gravity framework.
\end{abstract}

\section{Introduction}
\label{intro}
The concept of warp drives was first introduced by Miguel Alcubierre in his seminal 1994 paper~\cite{Alcubierre:1994tu}, and elaborated upon by José Natario in~\cite{Natario:2001tk}. A warp drive is a solution of General Relativity that has the appearance of a "bubble" propagating on some (flat or non--flat) spacetime background. The observers inside the bubble are in an inertial reference frame, which means warp drives do not require external energy sources to accelerate, and they may move at any speed (in principle including superluminal). This makes them a viable candidate for interstellar travel, but they have one significant downside: in order to sustain a bubble, one requires exotic matter with negative energy density.

As described by Homer Ellis in~\cite{Ellis:2004aw}, the Schwarzschild metric, which describes a black hole, can be mapped to a warp drive--type metric with the use of a coordinate system known as Painlevé--Gullstrand coordinates, which makes it possible to embed a warp drive in a black hole background. In section~\ref{sec:Correspondence} of this paper, we generalize this result to an arbitrary static spherically symmetric metric and show that it corresponds to a generalization of Alcubierre--Natario warp drives with non--flat intrinsic metric. We also demonstrate that warp drives can traverse Schwarzschild horizons without "feeling" them, while their embedding into more generic metrics without horizons produces a singularity. Even more crucially, we show that an external gravitational field can decrease the amount of negative energy required to sustain a warp drive, a fact which may be useful for the development of microscopic warp drives in a lab. Then, in section~\ref{sec:Analogue_Grav}, we suggest an analogue gravity framework that could also help to study the physics of warp drives going through horizons in lab experiments. We conclude in section~\ref{sec:conclusions} with an overview of these key proposals of the paper.
\section{The Black Hole--Warp Drive Correspondence}\label{sec:Correspondence}
A warp drive metric, as defined by Natario in~\cite{Natario:2001tk}, is given by
\begin{equation}
\label{Natario}
-dt^2+\sum^3_{i=1}(dx^i-\beta^i(\vec{r},t)dt)^2 \ .
\end{equation}
In ADM variables, one can define it by setting the lapse function $N$ to $1$, the shift vector $N^i$ to $\beta^i$, and the inner metric $h_{ij}$ to $\delta_{ij}$.

The warp drive itself is a localized perturbation of the metric moving on some (flat or non--flat) spacetime background. Assuming it's moving along the x--axis, its velocity is given by
\begin{equation}
\vec{v}_s(t)=\frac{dx_s}{dt} \ ,
\end{equation}
where $x_s(t)$ is the positon of the soliton's center.

This means that the functions $\beta^i$ have the form
\begin{equation}
\beta^i \ = \ (1-f(r_s))\beta^i_{out}(\vec{r},t)+f(r_s)\beta^i_{in}(t) \ ,
\end{equation}
where $\beta^i_{out}(\vec{r},t)$ is the background metric, $\beta^i_{in}(t)$ is the perturbation, and $f(\vec{r}_s(t))$ is a bell--shaped function describing the shape of the warp, with $r_s$ given by
\begin{equation}
r_s(t) \ = \ \sqrt{(x-x_s(t))^2+y^2+z^2} \ .
\end{equation}
In the particular case when the background metric is spherically symmetric, $\beta^i_{out}$ are given by
\begin{equation}
\beta^i_{out}(r,t) \ = \ \beta(r,t)\frac{x^i}{r} \ .
\end{equation}

In this case, the background metric can also be written in the more compact form in spherical coordinates
\begin{equation}\label{C1}
-dt^2+(dr-\beta(r,t)dt)^2+r^2d\Omega^2 \ .
\end{equation}
As shown by Painlevé and Gullstrand, the Schwarzshild metric
\begin{equation}
-(1-\frac{2GM}{r})dt^2+\frac{dr^2}{1-\frac{2GM}{r}}+r^2d\Omega^2
\end{equation}
can be brought to the form~\eqref{C1} with 
\begin{equation}
\beta \ = \ -\sqrt{\frac{2GM}{r}}
\end{equation}
via a coordinate transformation
\begin{equation}\label{PG}
t \ = \ T \ - \ \int \ dr \ \frac{\sqrt{\frac{2GM}{r}}}{1-\frac{2GM}{r}} \ .
\end{equation}
As suggested by Ellis in~\cite{Ellis:2004aw}, this relation can be used to embed an actual warp drive within the exterior of a black hole by replacing
\begin{equation}\label{C2}
\beta^i \ = \ \beta\frac{x^i}{r} \rightarrow (1-f(r_s))\beta\frac{x^i}{r}+f(r_s)\beta^i_{in}(t) \ .
\end{equation}
In this paper, we are going to generalize the Ellis' approach for every spherically symmetric metric, whose line element is given by
\begin{equation}
-e^{2\Phi(r)}dt^2 \ + \ \frac{dr^2}{1-\frac{b(r)}{r}} \ + \ r^2d\Omega^2 \ .  \label{metric}
\end{equation}
This metric is sufficiently general to represent a large variety of cases. We need also to include the following additional condition%
\begin{equation}
b(r_{0})\ =\ r_{0},\qquad r\in \left[ \ r_{0},\infty \right).   \label{throat}
\end{equation}%
A generic coordinate transformation has the form
\begin{equation}
dt \ = \ \xi dT \ + \ \eta dr \ ;
\end{equation}
however, since our metric background is time independent, then, given the condition 
\begin{equation}
\partial_r\xi \ = \ \partial_T\eta \ ,
\end{equation}
$\xi$ has to be a constant that can be set to $1$ by rescaling.

The transformation we seek is therefore given by
\begin{equation}\label{T}
t \ = \ T \ - \ \int \ dr \ \frac{\sqrt{e^{-2\Phi(r)}-1}}{\sqrt{1-\frac{b(r)}{r}}} \ ,
\end{equation}
and it allows to represent the line element as
\begin{equation}\label{E}
-dT^2+(g(r)dr-\beta(r)dT)^2+r^2d\Omega^2 \ ,
\end{equation}
with 
\begin{equation}\label{B}
\beta \ = \ -\sqrt{1-e^{2\Phi(r)}} \ .
\end{equation}
It differs from the standard Natario--type metric~\eqref{C1} by the form factor
\begin{equation}
g(r)\ =\ \frac{e^{\Phi \left( r\right) }}{\sqrt{1-\frac{b\left( r\right) }{r}%
}}\ .
\end{equation}

Note that because of condition $\left( \ref{throat}\right) $, $g(r)$ has a
singularity in $r\ =\ r_{0}$. In Cartesian coordinates (which have to be used to embed the warp drive, as it breaks the spherical symmetry) this metric has the form
\begin{equation}\label{C3}
-dT^2+h_{ij}(dx^i+N^idT)(dx^j+N^jdT) \ ,
\end{equation}
with
\begin{equation}\label{Ni}
N^i=-\left(1-f(r_s)\right)\frac{\beta}{g}\frac{x^i}{r}-f(r_s)v^i_s(T) \ ,
\end{equation}
and intrinsic metric
\begin{eqnarray}
h_{ij} \ = \ \delta_{ij}+(g^2-1)\frac{x_ix_j}{r^2}\\ 
\left(h^{ij} \ = \ \delta_{ij}+(g^{-2}-1)\frac{x_ix_j}{r^2}\right) \ .
\end{eqnarray}
Since the lapse function $N=1$, the Eulerian observers with geodesics
\begin{equation}
u_\mu \ = \ (1,0,0,0)
\end{equation}
are still in free--fall, and therefore the metric~\eqref{C3} can also be considered a kind of warp drive metric, albeit different from the Natario--type ones (a warp drive with non--flat spatial slices has previously been proposed in~\cite{VanDenBroeck:1999sn}).

Because of the presence of $g(r)$, the induced metric $h_{ij}$ has a singularity too. Such
a singularity can be eliminated if $g(r)$ is finite at $r=r_{0}$. Since
\begin{equation}
e^{2\Phi} \ = \ g^2(r)\left(1-\frac{b(r)}{r}\right) \ ,
\end{equation}
this means that the line element $\left( \ref{metric}\right) $ would have a
horizon.

The simplest case
\begin{gather}
g(r)=1 \ , \ b(r)=const \label{Schw}
\end{gather}
corresponds to the Schwarzschild metric. Thanks to Eq.~\eqref{Schw}, the metric is flat and the 3D scalar curvature vanishes. Therefore the energy density can be expressed as
\begin{equation}
\rho \ = \ \frac{1}{16\pi G}\left(K^2-K_{ij}K^{ij}\right) \ ,
\end{equation}
with the extrinsic curvature tensor given by
\begin{equation}
K_{ij} \ = \ \frac{1}{2}\left(\partial_iN_j+\partial_jN_i\right)
\end{equation}
in Cartesian coordinates.

Now, let us assume that the warp drive is moving along the x--axis:
\begin{equation}
N^x \ = \ (1-f)\sqrt{\frac{R_G}{r}}\frac{x}{r}-fv \ ,
\end{equation}
and consider the limit in which the characteristic size of the warp drive (the support of the function $f$) is much smaller than the Schwarzschild radius of the black hole $R_G$. In this limit, $N_{y,z}\approx0$, and the energy density reduces to
\begin{equation}
\rho \ = \ -\frac{1}{32\pi G}\left((\partial_yN^x)^2+(\partial_zN^x)^2\right) \ ,
\end{equation}
from which, once again neglecting terms $\propto\frac{y}{r}$ and $\frac{z}{r}$, we obtain
\begin{equation}
-\frac{1}{32\pi G}\left(v+\sqrt{\frac{R_G}{r}}\right)^2f'^2\left(\frac{y^2+z^2}{r^2_s}\right)
\end{equation}
Hence, for negative $v$ (i. e. the warp drive moving towards the black hole) with
\begin{equation}
|v|>\frac{1}{2}\sqrt{\frac{R_G}{r}}
\end{equation}
\textit{the amount of negative energy required to sustain a warp drive would be decreased by the black hole's gravitational field} (otherwise, or if the warp drive is moving in the opposite direction, it would be increased).

As shown in~\cite{White:2021hip}, it may be possible to create warp drive--like structures within Casimir cavities in a lab, so it appears a promising direction to investigate how they would be affected by an external gravitational field (and whether such a field can serve a practical purpose by making it possible to realize a warp drive with a lesser amount of negative energy).

Finally, let us note that, as the $00$--component of the metric tensor is approximately
\begin{equation}
g_{00} \ \approx \ -1+\left((1-f)\sqrt{\frac{R_G}{r}}+f|v|\right)^2 \ ,
\end{equation}
\textit{the horizon would be effectively absent inside the warp bubble as long as it is subluminal} ($v<1$).
\section{Analogue Gravity Model}\label{sec:Analogue_Grav}
While warp drives remain at the moment beyond the realm of experiment, one could, at least in principle, try to gain some understanding of the underlying physics through an analogue gravity set--up. The framework of analogue gravity is rooted in the observation that for a barotropic, inviscid, irrotational fluid, the perturbations of the velocity potential $\phi$ obey the equation~\cite{Barcelo:2005fc}
\begin{equation}
\frac{1}{\sqrt{-g}}\partial_\mu(\sqrt{-g}g^{\mu\nu}\partial_\nu\phi) \ = \ 0 \ ,
\end{equation}
which is the equivalent of the Klein--Gordon equation with the effective metric
\begin{equation}\label{C5}
g_{\mu\nu} \ = \ \frac{\rho_0}{c_s} \begin{pmatrix}
-(c_s^2-v^2) & \vec{v}^T \\
\vec{v} & \delta_{ij}
\end{pmatrix} \ .
\end{equation}
Here $c_s$ is the speed of sound in the medium, given by
\begin{equation}
c^2_s \ = \ (\frac{\partial\rho}{\partial p})^{-1} \ ,
\end{equation}
and $\vec{v}$ is the background velocity of the fluid. For a radially symmetric system, it may also be written in the form
\begin{equation}
\frac{\rho_0}{c_s}\left(-(c^2_s-v^2)dt^2+2vdtdr+dr^2+r^2d\Omega^2\right) \ .
\end{equation}
In particular, the Gross--Pitaevskii equation
\begin{equation}
i\partial_t\psi \ = \ -\frac{\hbar^2}{2m}\triangle\psi+g|\psi|^2\psi \ ,
\end{equation}
which describes a Bose--Einstein condensate, can be written in the hydrodynamic form in the Madelung representation, so that the phase of the wave function would play the role of the velocity potential, and the speed of sound would be
\begin{equation}
c^2_s \ = \ \frac{g}{m}n_c \ ,
\end{equation}
where $n_c$ is the background density of the condensate, given by $|\psi|^2$.

The metric~\eqref{C5} is, up to the prefactor $\frac{\rho}{c_s}$, exactly the warp drive metric~\eqref{Natario}, and hence one could imitate a warp drive by introducing a perturbation in the velocity profile, as suggested in~\cite{Fischer:2002jn}~\cite{Finazzi}. By performing the Painlevé--Gullstrand transformation~\eqref{PG} in reverse, one can map this metric to
\begin{equation}\label{C6}
\frac{\rho_0}{c_s}\left(-(c^2_s-v^2)dt^2+\frac{c_s^2}{c_s^2-v^2}dr^2+r^2d\Omega^2\right) \ ,
\end{equation}
which is exactly the Schwarzschild metric for the radial velocity profile
\begin{equation}
v \ = \ c_s\sqrt{\frac{2GM}{r}} \ .
\end{equation}
A 1+1--dimensional version of this setup was realized experimentally with a Bose--Einstein condensate of Rb atoms~\cite{Lahav:2009wx}, making it possible to observe the analogue Hawking radiation~\cite{Steinhauer:2015saa}. In principle, one can combine this setup with the one described in~\cite{Finazzi} to realize a condensate velocity profile given by
\begin{equation}\label{C7}
v(t,x) \ = \ v_0(x)+\delta v(x-Vt) \ ,
\end{equation}
where $v_0$ is the background profile that has a step--like transition from subsonic to supersonic at some point $x_0$, and $\delta v$ is a solitonic wave in the velocity profile propagating at velocity $V$. By having $\delta v$ cross the point $x_0$, one can simulate the scenario described in~\cite{Ellis:2004aw}, in which a warp drive crosses a black hole horizon.

The more generic class of metrics of $g(r)\neq1$ may be simulated by considering a variable speed of sound: in particular, it has been suggested in~\cite{Mateos:2017sjv} that one could modulate the speed of sound using the Feshbach resonance. In this case, we can make the identifications
\begin{eqnarray}
c_s(r) \ = \ g(r) \ = \ \frac{e^{\Phi(r)}}{\sqrt{1-\frac{b(r)}{r}}} \ ,\\
v(r) \ = \ g(r)\sqrt{\frac{b(r)}{r}}
\end{eqnarray}
to obtain from~\eqref{C6} a metric equivalent to~\eqref{metric}, up to the conformal factor $\frac{\rho_0}{c_s}$ in front. In principle, it can be set to a constant if one chooses a fluid with the barotropic equation of state
\begin{equation}
\rho \ \propto \ p^{1/3} \ .
\end{equation}
By combining this setup with~\eqref{C7}, it may be possible to simulate the passage of a warp drive through the horizon.

\section{Conclusions}\label{sec:conclusions}
In this paper, we studied a class of coordinate transformations that generalizes Painlevé--Gullstrand coordinates to an arbitrary spherically symmetric metric and makes it possible to embed a warp drive into a given metric background. We found that the presence of a horizon is required to avoid a singularity. For the specific case of Schwarzschild black hole, the horizon would be effectively absent inside the warp bubble as long as it's subluminal, making the warp drive a possible safe route for crossing horizons.

More importantly, we discovered that external gravitational field can decrease the amount of negative energy required to sustain a warp drive, which may be instrumental for creating microscopic warp drives with Casimir cavities, such as the one described in~\cite{White:2021hip}.

In the second part of the paper, we proposed another possible application of our results for tabletop experiments in the form of an analogue gravity set--up. For this purpose, we started with a known realization of analogue black holes with Bose--Einstein condensate and supplemented it with a warp drive analogue in the form of a solitonic wave. We also suggested that in order to incorporate the more generic non--Schwarzschild spherically symmetric metrics, one needs to introduce spatial variations in the condensate speed of sound.

In~\cite{GarattiniZatrimaylov} we shall explore the tangential topic of warp drive metrics with intrinsic curvature, generalizing the results of~\cite{VanDenBroeck:1999sn}. This class of models can possibly reduce the amount of exotic matter with negative energy density required to build a warp drive.

\begin{acknowledgements}
We are grateful to Harold "Sonny" White for the discussions and for his useful feedback on earlier versions of this work, and to Prof. Claudio Maccone for his instructive comments and questions. The work is supported by the 2023 LSI grant "Traversable Wormholes: A Road to Interstellar Exploration". Part of the computations in this work was done with OGRe, a General Relativity Mathematica package developed by Barak Shoshany~\cite{Shoshany:2021iuc}.
\end{acknowledgements}

\appendix
\section{Generic warp drives}
We may also consider the generic case when the warp drive is not necessarily much smaller than the black hole, and not necessarily moving in the radial direction. Following~\cite{Santiago:2021aup}, one can write the energy density in the form
\begin{equation}\label{rho}
\rho \ = \ \frac{1}{16\pi G}\left(\partial_i(N_i\partial_jN_j-N_j\partial_jN_i)-\frac{1}{4}(\partial_iN_j-\partial_jN_i)^2\right) \ .
\end{equation}
Now, if we take
\begin{equation}
N^i \ = \ (1-f)\sqrt{\frac{R_G}{r}}\frac{x^i}{r}-fv^i \ ,
\end{equation}
the first term vanishes, and the second term becomes
\begin{equation}
\frac{f'^2}{32\pi Gr^2_s}\left|\left(\vec{v}+\sqrt{\frac{R_G}{r}}\frac{\vec{r}}{r}\right)\otimes r_s\right|^2 \ = \ \frac{f'^2\sin^2\theta}{32\pi G}\left|\vec{v}+\sqrt{\frac{R_G}{r}}\frac{\vec{r}}{r}\right|^2 \ .
\end{equation}
The condition for the modulus of this expression to be decreased is
\begin{equation}
v^2+2v\sqrt{\frac{R_G}{r}}\cos\psi+\frac{R_G}{r} \ < \ v^2 \ ,
\end{equation}
where $\psi$ is the angle between the vectors $\vec{r}$ and $\vec{v}$. From this we get the condition
\begin{equation}
v\cos\psi \ < -\frac{1}{2}\sqrt{\frac{R_G}{r}} \ ,
\end{equation}
i. e. the projection of $\vec{v}$ on the $r$--axis should be negative and greater by modulus than $\frac{1}{2}\sqrt{\frac{R_G}{r}}$. From this we automatically get the weaker condition
\begin{equation}
|v| \ > \frac{1}{2}\sqrt{\frac{R_G}{r}} \ .
\end{equation}
\end{document}